\documentclass[aps,pra,superscriptaddress,amsmath,amssymb,preprintnumbers,showpacs]{revtex4}
\usepackage{amssymb}
\usepackage{graphicx}
\usepackage{bm}
\usepackage{color}

\newcommand{\Ignore}[1]{}

\newcommand{\Ket}[1]{\left\vert #1\right\rangle}
\newcommand{\Bra}[1]{\left\langle #1\right\vert}

\newcommand{\KetBra}[2]{\left\vert#1\right\rangle\left\langle#2\right\vert}
\newcommand{\Projector}[1]{\KetBra{#1}{#1}}

\renewcommand{\eqref}[1]{(\ref{#1})} 

\def\eff{\text{eff}}
\def\phen{\text{phen}}
\def\bare{\text{bare}}

\begin{document}
\title{Stimulated Raman adiabatic passage in an open quantum system: Master equation approach}

\author{M. Scala}
\affiliation{Dipartimento di Scienze Fisiche ed Astronomiche dell'Universit\`{a} di Palermo, Via Archirafi 36, 90123 Palermo, Italy}

\author{B. Militello}
\affiliation{Dipartimento di Scienze Fisiche ed Astronomiche dell'Universit\`{a} di Palermo, Via Archirafi 36, 90123 Palermo, Italy}

\author{A. Messina}
\affiliation{Dipartimento di Scienze Fisiche ed Astronomiche dell'Universit\`{a} di Palermo, Via Archirafi 36, 90123 Palermo, Italy}

\author{N. V. Vitanov}
\affiliation{Department of Physics, Sofia University, James Bourchier 5 blvd, 1164 Sofia, Bulgaria}
\affiliation{Institute of Solid State Physics, Bulgarian Academy of Sciences, Tsarigradsko chauss\'{e}e 72, 1784 Sofia, Bulgaria}

\begin{abstract}
A master equation approach to the study of environmental effects
in the adiabatic population transfer in three-state systems is
presented. A systematic comparison with the non-Hermitian
Hamiltonian approach [N. V. Vitanov and S. Stenholm, Phys. Rev. A
{\bf 56}, 1463 (1997)] shows that in the weak coupling limit the
two treatments lead to essentially the same results. Instead, in
the strong damping limit the predictions are quite different: in
particular the counterintuitive sequences in the STIRAP scheme
turn out to be much more efficient than expected before. This
point is explained in terms of quantum Zeno dynamics.
\end{abstract}


\maketitle


\section{Introduction}

Stimulated Raman adiabatic passage (STIRAP) is a powerful
technique for coherent population transfer in a three-state
chainwise connected system 1-2-3 \cite{STIRAP original 1, STIRAP
original 2, STIRAP review AAMOP, STIRAP review ARPC, STIRAP review
RMP 1, STIRAP review RMP 2}. In the most common $\Lambda$ linkage
pattern, states $\Ket{1}$ and $\Ket{3}$ are ground or metastable
levels, while the intermediate state $\Ket{2}$ is a decaying
excited electronic level. The unique advantage of STIRAP over
other population transfer techniques is that in the adiabatic
limit this intermediate state $\Ket{2}$ does not receive even
transient population during the transition $\Ket{1}\to \Ket{3}$.
This feature derives from the fact that STIRAP proceeds via a dark
state, which is a superposition of states $\Ket{1}$ and $\Ket{3}$
only. A two-photon resonance between states $\Ket{1}$ and
$\Ket{3}$ ensures the emergence of such an eigenstate of the
Hamiltonian. A counterintuitive pulse sequence, Stokes-pump (with
the Stokes driving the 2-3 transition and the pump driving the 1-2
transition), aligns initially state $\Ket{1}$ with the dark state,
Finally, adiabatic evolution, which is enforced by selecting
sufficiently large pulse areas, ensures that the three-level
system remains in the dark state at all times until it aligns with
the target state $\Ket{3}$ in the end.

In the adiabatic limit, the properties of the intermediate state $\Ket{2}$, including its detuning and loss rates (e.g., spontaneous
emission within and outside the system), are irrelevant because it
is decoupled from the dynamics. However, these factors cannot be
ignored completely because first, in a real experiment the
evolution is never perfectly adiabatic, and second, these factors
affect the adiabatic condition itself. The robustness of STIRAP
against the intermediate-level detuning  has been quantified in
\cite{ref:Vitanov1997b}. The effects of dephasing \cite{STIRAP
dephasing} and spontaneous emission within the system \cite{STIRAP
spontaneous emission} have also been scrutinized.

The effect of irreversible population loss from the intermediate
state $\Ket{2}$ has been studied in \cite{ref:Vitanov1997} wherein the
decay has been introduced \emph{phenomenologically}, adding an
imaginary diagonal term in the lossless Hamiltonian. Nevertheless,
because of the rapidly increasing popularity of STIRAP as a
quantum control tool in dissipative environments, it is interesting, instructive and important to treat the
problem with greater mathematical rigor, starting from a
microscopic model, which explicitly takes into account the coupling
between the system and an external environment.

For time-independent models, there are well established techniques
which allow for a description of the open dynamics of the system
of interest by means of master equations, which can be
systematically derived from the system Hamiltonian
\cite{ref:Gardiner,ref:Petru}. However, for time-dependent
Hamiltonian models the derivation of the master equation requires
more attention. A fully satisfactory and very simple theory of
master equations for such systems has been developed by Davies in
the late 70's \cite{ref:Davies1978}. The main feature of this
approach is that, under the hypothesis of very short reservoir
correlation times, one obtains a time-dependent master equation
describing jumps between instantaneous eigenstates of the system
Hamiltonian. This approach has been used recently in the study of
quantum logic gates based on the accumulation of geometric phases
in adiabatic evolutions \cite{ref:Florio2006}. Similar
time-dependent master equations have been used by other authors in
problems involving quantum adiabatic evolution
\cite{ref:Carollo,ref:Lidar}.

We emphasize  that in general the microscopic derivation of a
master equation for a given physical system may give rise to
predictions which differ from the ones obtained from
phenomenological models, as recently seen for example  in the
context of lossy cavity QED \cite{ref:Turku-1,ref:Turku-2,ref:Czachor-1,ref:Czachor-2} and two-qubit dynamics \cite{ref:JPAScala}.

In this paper, we present a microscopic model from which we derive
a master equation describing the dissipative dynamics of a system
subjected to a STIRAP scheme. In order to perform a systematic
comparison between our model and the phenomenological model in
\cite{ref:Vitanov1997}, we move to a description of the dynamics
by means of an effective non-Hermitian Hamiltonian, in this case
equivalent to the master equation approach. Comparing the
predictions coming from the two non-Hermitian Hamiltonian models,
we find that, according to our effective model, the population
transfer is more efficient than previously expected. The
discrepancy is more evident in the limit of strong damping, where
the new effects found can be easily understood in terms of quantum
Zeno dynamics.

The paper is structured as follows. In the next section we recall
the main properties of the system under scrutiny which has been
described in \cite{ref:Vitanov1997} and derive the master equation
starting from a microscopic model of system-reservoir coupling. In
the third section we derive the relevant effective Hamiltonian,
while in the fourth section we compare the predictions from the
effective and the phenomenological Hamiltonian models. Finally, in
the last section some conclusive remarks are given.


\section{Master Equation}

\begin{figure}[tb]
\includegraphics[width=0.30\textwidth, angle=90]{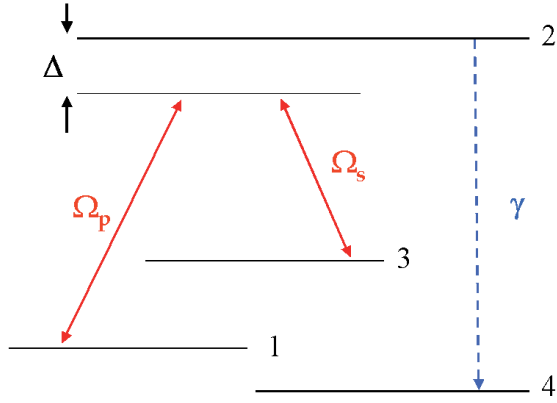}
\caption{(Color online) Level scheme and coupling scheme. Level
$4$ is very far from the other three.} \label{fig:Couplings}
\end{figure}

We model the irreversible population decay from the intermediate
state  $\Ket{2}$ by introducing an additional state $\Ket{4}$, to
which state $\Ket{2}$ decays (see figure \ref{fig:Couplings}). In
the absence of such a decay the Hamiltonian describing the
enlarged four-state system reads (with $\hbar=1$):
\begin{widetext}
\begin{equation}
H_0(t) = \left[\begin{array}{cccc}
\omega_1 & \Omega_p(t) e^{i(\omega_{21}-\Delta)t} & 0 & 0 \\
\Omega_p(t)e^{-i(\omega_{21}-\Delta)t} & \omega_2 & \Omega_s(t)e^{-i(\omega_{23}-\Delta)t} & 0 \\
0 & \Omega_s(t)e^{i(\omega_{23}-\Delta)t} & \omega_3 & 0 \\
0 & 0 & 0 & \omega_4 \\
\end{array}\right]\,,
\end{equation}
\end{widetext}
with $\omega_4<0$ well separated from the other levels, and
$\omega_{mn}=\omega_m-\omega_n$. The population decay of state
$\Ket{2}$ to state $\Ket{4}$ is modeled by a system-bath coupling
Hamiltonian involving these two states, which in the non-rotating
frame is given by:
\begin{equation}
H_{\text{dec}} = \left(\Ket{2}\Bra{4}+\Ket{4}\Bra{2}\right)\otimes\sum_{k}g_k\left(a_k+a_k^\dag\right)\,.
\end{equation}
In the rotating frame, the transformation to which is given by
$T(t)=e^{i\omega_{1}t}\Projector{1}+e^{i(\omega_{2}-\Delta)t}\Projector{2}+e^{i\omega_{3}t}\Projector{3}$,
the Hamiltonian of the system becomes
\begin{equation}
H_s(t)= \left[\begin{array}{cccc}
0 & \Omega_p(t) & 0 & 0 \\
\Omega_p(t) & \Delta & \Omega_s(t) & 0 \\
0 & \Omega_s(t) & 0 & 0 \\
0 & 0 & 0 & \omega_4 \\
\end{array}\right]\,,
\end{equation}
and $H_{\text{dec}}$ changes accordingly:
\begin{eqnarray}\label{eq:Hdec}
H_{\text{dec}} &=& \nonumber
\left[e^{i(\omega_{2}-\Delta)t}\Ket{2}\Bra{4}+e^{-i(\omega_{2}-\Delta)t}\Ket{4}\Bra{2}\right] \\
&\otimes&\sum_{k}g_k\left(a_k+a_k^\dag\right)\,.
\end{eqnarray}

The instantaneous eigenstates of the Hamiltonian $H_s(t)$ are
\cite{ref:Vitanov1997}:
\begin{subequations}
\begin{align}
\Ket{+}&= \sin\varphi\sin\theta\Ket{1}+\cos\varphi\Ket{2}+\sin\varphi\cos\theta\Ket{3}\,,\\
\Ket{0}&= \cos\theta\Ket{1}-\sin\theta\Ket{3}\,,\\
\Ket{-}&=
\cos\varphi\sin\theta\Ket{1}-\sin\varphi\Ket{2}+\cos\varphi\cos\theta\Ket{3}\,,
\end{align}
\end{subequations}
where
\begin{subequations}
\begin{align}
\tan\theta(t)&=\frac{\Omega_p(t)}{\Omega_s(t)}\,,\\
\tan{2\varphi(t)}&=\frac{2\Omega_0(t)}{\Delta(t)}\,,\\
\Omega_0(t)&=\sqrt{\Omega_p(t)^2+\Omega_s(t)^2}\,,
\end{align}
\end{subequations}
and state $\Ket{4}$ which is left unchanged by the transformation.
The corresponding eigenvalues are $\omega_+=\Omega_0 \cot\varphi$,
$0$, $\omega_-=-\Omega_0\tan\varphi$ and $\omega_4$.

In the ideal case (in the absence of $H_{\text{dec}}$), perfect
population transfer from $\Ket{1}$ to $\Ket{3}$ takes place. In
particular, two different pulse sequences are possible, the
\emph{intuitive} and the \emph{counterintuitive} sequences. In the
intuitive sequence, the pump pulse $\Omega_p(t)$ precedes the
Stokes pulse $\Omega_s(t)$; then for nonzero single-photon
detuning $\Delta$ it can be shown that in the adiabatic limit the
population remains at all times in state $\Ket{-}$, which at
$t=-\infty$ is equal to state $\Ket{1}$ and at $t=\infty$ it is
equal to state $\Ket{3}$. Therefore, in the adiabatic limit there
is a perfect population transfer from $\Ket{1}$ to $\Ket{3}$
\cite{ref:Vitanov97 analytic}; this process has recently been
termed b-STIRAP (because it proceeds via the \emph{bright} state
$\Ket{-}$) \cite{ref:b-STIRAP}. For the counterintuitive pulse
sequence, the pump pulse follows the Stokes pulse. Then the
population is transferred through the \emph{dark} state $\Ket{0}$,
which again is equal to $\Ket{1}$ at $t=-\infty$ and to $\Ket{3}$
at $t=\infty$.

In the non-ideal case of decaying state $\Ket{2}$, the predictions
change. In ref. \cite{ref:Vitanov1997} it has been shown that the
instability of the intermediate state reduces the efficiency of
the scheme. In particular it turns out that the intuitive sequence
is more fragile than the counterintuitive one. Since these results
have been obtained by means of a phenomenological Hamiltonian
approach, it is natural to ask whether these results may change
when a microscopic master equation approach is used. In the
following we give the master equation describing the dissipative
dynamics of the system.

According to the general theory by Davies \cite{ref:Davies1978},
recently applied to some time-dependent systems
\cite{ref:Florio2006}, under the assumption of bath correlation
times much smaller than the time of variation of the Hamiltonian,
the master equation describes jumps between the instantaneous
eigenstates of the time-dependent Hamiltonian. Since the
Hamiltonian (\ref{eq:Hdec}) involves only states $\Ket{2}$ and
$\Ket{4}$, and since state $\Ket{2}$ is not involved in the dark
state $\Ket{0}$, the only jumps allowed are
$\Ket{+}\leftrightarrows\Ket{4}$ and
$\Ket{-}\leftrightarrows\Ket{4}$. Therefore the master equation of
our system is given by
\begin{eqnarray}
\nonumber
\frac{\mathrm{d}\rho}{\mathrm{d}t}&=&-i[H_s(t),\rho]\\
\nonumber
&+&\gamma_+(t)\left(\Ket{4}\Bra{+}\rho\Ket{+}\Bra{4}-\tfrac{1}{2}\left\{\Projector{+}, \rho\right\}\right)\\
\nonumber
&+&\gamma_-(t)\left(\Ket{4}\Bra{-}\rho\Ket{-}\Bra{4}-\tfrac{1}{2}\left\{\Projector{-}, \rho\right\}\right)\\
\nonumber
&+&\gamma_+^{exc}(t)\left(\Ket{+}\Bra{4}\rho\Ket{4}\Bra{+}-\tfrac{1}{2}\left\{\Projector{4}, \rho\right\}\right)\\
&+&\gamma_-^{exc}(t)\left(\Ket{-}\Bra{4}\rho\Ket{4}\Bra{-}-\tfrac{1}{2}\left\{\Projector{4},
\rho\right\}\right)\,.\label{eq:master_equation}
\end{eqnarray}
Here
\begin{subequations}\label{eq:DefGamma}
\begin{align}
\gamma_+(t)&=\cos^2\varphi(t)|g(\tilde{\omega}_{+})|^2 D(\tilde{\omega}_{+})(N(\tilde{\omega}_{+})+1)\,,\\
\gamma_-(t)&=\sin^2\varphi(t)|g(\tilde{\omega}_{-})|^2 D(\tilde{\omega}_{-})(N(\tilde{\omega}_{-})+1)\,,
\end{align}
\end{subequations}
are the decay rates of the states $\Ket{+}$ and $\Ket{-}$ toward
state $\Ket{4}$, corresponding to the Bohr frequencies
$\tilde{\omega}_\pm=\omega_\pm+\omega_2-\omega_4$. The quantities
$D(\tilde{\omega}_{\pm})$ and $N(\tilde{\omega}_{\pm})$ are
respectively the densities of modes and the average numbers of
photons in the reservoir at the relevant frequencies. The
parameter $g(\tilde{\omega}_{\pm})$ gives the continuum-limit
system-reservoir coupling strengths. We note that the factors
$\cos^2\varphi$ and $\sin^2\varphi$ in Eq.~\eqref{eq:DefGamma}
come from the calculation of the square moduli of the matrix
elements of the system operator $\Ket{2}\Bra{4}+\Ket{4}\Bra{2}$
appearing in $H_{\text{dec}}$ between states $\Ket{4}$ and
$\Ket{+}$, and between $\Ket{4}$ and $\Ket{-}$, respectively.
Finally, the excitation rates are given by
$\gamma_\pm^{exc}(t)=\gamma_\pm(t)
N(\tilde{\omega}_{\pm})/(N(\tilde{\omega}_{\pm})+1)$ and vanish at
zero temperature. This will be the case in the following sections.

\section{Rate Equations and Effective Hamiltonian}

Introducing the density matrix decomposition in terms of the instantaneous eigenstates of the Hamiltonian $H_s(t)$,
\begin{equation}\label{eq:rho_expansion}
\rho(t)=\sum_{ij} \rho_{ij}(t)\KetBra{i(t)}{j(t)},
\end{equation}
and substituting it into the master equation (\ref{eq:master_equation}), one obtains a set of rate equations,
\begin{equation}
\dot\rho_{ij}(t)=\sum_k A_{ijkl}(t)\rho_{kl}(t)\,.
\end{equation}
On the other hand, introducing the matrix of the density operator matrix elements in the time-dependent basis, i.e.
\begin{equation}
\bar{\rho}=\left[\rho_{ij}(t)\right]\,,
\end{equation}
with $i,j=\pm,0$, one can prove (see appendix \ref{AppA}) that the
same rate equations can be obtained with a pseudo-Liouville
equation, restricted to  the subspace $\Ket{+}$, $\Ket{0}$,
$\Ket{-}$:
\begin{equation}\label{eq:Pseudo_Liouville}
\dot{\bar{\rho}}=-i \left( H_{\eff}\bar{\rho}-\bar{\rho}
H_{\eff}^\dag \right)\,,
\end{equation}
with
\begin{widetext}
\begin{equation}\label{eq:Effective_Hamiltonian}
H_{\eff}=\left[\begin{array}{cccc}
\Omega_0\cot\varphi -i\Gamma\cos^2\varphi & i\dot{\theta}\sin\varphi & i\dot{\varphi}\\
-i\dot{\theta}\sin\varphi & 0 & -i\dot{\theta}\cos\varphi\\
-i\dot{\varphi} & i\dot{\theta}\cos\varphi & -\Omega_0\tan\varphi-i\Gamma\sin^2\varphi\\
\end{array}\right]\,,
\end{equation}
where, reminding that $N(\tilde{\omega}_{\pm})=0$ at zero temperature and assuming flat reservoir spectrum \cite{ref:Petru},
we have set $\Gamma=|g(\tilde{\omega}_{+})|^2 D(\tilde{\omega}_{+})=|g(\tilde{\omega}_{-})|^2 D(\tilde{\omega}_{-})$.

The comparison of this Hamiltonian with the phenomenological Hamiltonian of Eq.~(4) of \cite{ref:Vitanov1997}, i.e.:
\begin{equation}\label{eq:Pseudo_Liouville-phen}
\dot{\bar{\rho}}=-i \left( H_{\phen}\bar{\rho}-\bar{\rho}
H_{\phen}^\dag \right)\,,
\end{equation}
where
\begin{equation}\label{H phenomenological}
H_{\text{phen}}=\left[\begin{array}{cccc}
\Omega_0\cot\varphi -i\Gamma\cos^2\varphi & i\dot{\theta}\sin\varphi & i\dot{\varphi}+\frac{i}{2}\Gamma\sin 2\varphi\\
-i\dot{\theta}\sin\varphi & 0 & -i\dot{\theta}\cos\varphi\\
-i\dot{\varphi}+\frac{i}{2}\Gamma\sin 2\varphi & i\dot{\theta}\cos\varphi & -\Omega_0\tan\varphi-i\Gamma\sin^2\varphi\\
\end{array}\right]\,,
\end{equation}
\end{widetext}
shows that the only difference is in terms (1,3) and (3,1). The
two Hamiltonians \eqref{eq:Effective_Hamiltonian} and \eqref{H
phenomenological} are written down in the basis $\Ket{+}$,
$\Ket{0}$, $\Ket{-}$ (not $\Ket{1}$, $\Ket{2}$, $\Ket{3}$).

More details of the derivation of the effective Hamiltonian are given in Appendix \ref{AppA}.

\section{Comparison of the two approaches}

We shall now compare the two models, the one using the effective
Hamiltonian (\ref{eq:Effective_Hamiltonian}) derived from the
master equation, and the phenomenological one using the
Hamiltonian (\ref{H phenomenological}). We shall consider, as in
\cite{ref:Vitanov1997}, two pulses of the form
\begin{subequations}\label{eq:pulses}
\begin{equation}
\Omega_1(t)=\frac{\alpha}{T\sqrt{2}}\,\mathrm{sech}\left(\frac{t}{T}\right)\cos\left[\frac{\pi}{4}\left(\tanh\frac{t}{T}+1\right)\right]\,,
\end{equation}
\begin{equation}
\Omega_2(t)=\frac{\alpha}{T\sqrt{2}}\,\mathrm{sech}\left(\frac{t}{T}\right)\sin\left[\frac{\pi}{4}\left(\tanh\frac{t}{T}+1\right)\right]\,,
\end{equation}
\end{subequations}
where $T$ determines the widths of the pulses while $\alpha$ their
intensities, so that $\alpha T$ gives a measure of each pulse
area.

\subsection{Weak damping ($\Gamma T \ll 1$)}

\textbf{Intuitive pulse sequence --} For the intuitive sequence (i.e., when $\Omega_p=\Omega_1$ and $\Omega_s=\Omega_2$), the effective and phenomenological models deliver essentially the same results.
This feature derives from the fact that the adiabatically transferred population is that of state $\Ket{-}$.
In the weak-damping limit, the dominant source of decay is the population decay of $\Ket{-}$ itself due to the term (3,3) in both the Hamiltonians $H_{\eff}$ and $H_{\phen}$, thereby diminishing the
effects of the off-diagonal terms and in particular of $\frac{i}{2}\Gamma\sin 2\varphi$ in (1,3) and (3,1), which are the only difference between the two models.
Hence the decay of the final state, with a good approximation, is given by
\begin{equation}
P_3\simeq \exp\left[-2\Gamma\int_{-\infty}^\infty \sin^2 \varphi(t) \text{d} t \right].
\end{equation}

\begin{figure}[tb]
\begin{tabular}{c}
\includegraphics[width=0.45\textwidth, angle=0]{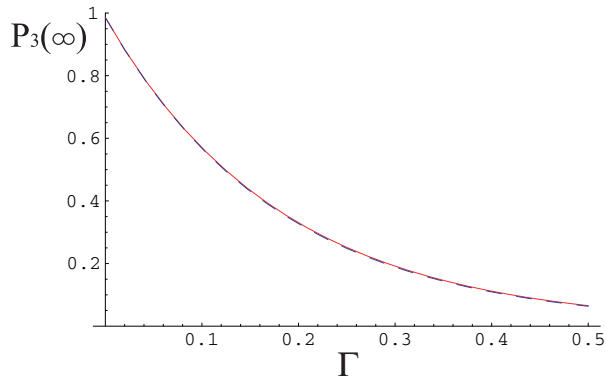}
\end{tabular}
\caption{(Color online) Asymptotic values of population $P_3$ for the phenomenological and effective (perfectly superimposed) model
for the intuitive sequence, as functions of $\Gamma$ (in units of $T^{-1}$), for $\alpha T=10$, $\Delta T=1$.}
\label{fig:Asympt_Intuitive}
\end{figure}

Figure \ref{fig:Asympt_Intuitive} shows the asymptotic value of $P_3$ as a function of $\Gamma$ for both models, and shows the
nearly perfect agreement of the two predictions, in particular the exponential dependence on $\Gamma$.

\textbf{Counterintuitive pulse sequence --} For the counterintuitive sequence ($\Omega_s=\Omega_1$ and $\Omega_p=\Omega_2$), still under the weak damping conditions,
 we can study the evolution through the adiabatic elimination, assuming that the states $\Ket{+}$ and $\Ket{-}$ are not very much populated during the process.
We decompose the state vector in the adiabatic basis as $\Ket{\psi}=a_+\Ket{+}+a_0\Ket{0}+a_-\Ket{-}$,
 and the Schr\"odinger equation becomes a set of linear differential equations for the probability amplitudes, 
\begin{equation}
i \frac{\text{d}}{\text{d}t}\left[\begin{array}{c} a_+\\ a_0\\ a_- \end{array}\right] = H\,
\left[\begin{array}{c} a_+\\ a_0\\ a_- \end{array}\right],
\end{equation}
with $H=H_{\eff}$ or $H=H_{\phen}$, depending on the model considered. By setting $\dot{a}_+\approx\dot{a}_-\approx 0$, we find $a_+$ and $a_-$ as functions of $a_0$,
 and substitute these expressions in the equation for $\dot{a}_0$, which assumes the form
\begin{equation}\label{eq:AdiabaticEqForA0}
\dot{a}_0(t) = - A(t) a_0(t)\,,
\end{equation}
where $A(t)$ has one of the following expressions depending on the model one is considering:
\begin{widetext}
%
\begin{equation}
A^{\phen}(t) =  \frac{\dot{\theta}^2 ( \Gamma +
2i\Omega_0\cot2\varphi)}
 {(i\Omega_0\cot\varphi+\Gamma\cos^2\varphi)(-i\Omega_0\tan\varphi+\Gamma\sin^2\varphi) + \dot{\varphi}^2-\frac14\Gamma^2\sin^2 2\varphi}
\end{equation}
%
%
for the phenomenological model, and
\begin{equation}
A^{\eff}(t) =  \frac{\dot{\theta}^2 [\Gamma(\cos^4\varphi +
\sin^4\varphi) + 2i\Omega_0\cot2\varphi ]}
{(i\Omega_0\cot\varphi+\Gamma\cos^2\varphi)(-i\Omega_0\tan\varphi+\Gamma\sin^2\varphi)
+ \dot{\varphi}^2}
\end{equation}
\end{widetext}
for the effective model.

Equation \eqref{eq:AdiabaticEqForA0} has the very simple solution
\begin{equation}
a_0(t) = a_0(-\infty) \exp\left[-\int_{-\infty}^t
A(t')\mathrm{d}t'\right]\,,
\end{equation}
which corresponds to the dark-state population
\begin{equation}
P_0(t) = \exp\left[-2\int_{-\infty}^t \Re(A(t'))\mathrm{d}t'\right]\,.
\end{equation}

For the phenomenological model, according to \cite{ref:Vitanov1997}, at the first order in the parameter $\Gamma$, we find
\begin{equation}
P_3^{\phen}(\infty) = 
 \exp\left[-2\Gamma\int_{-\infty}^\infty \frac{\dot\theta^2}{\Omega_0^2+\dot\varphi^2}\,\mathrm{d}t'\right]\,,
\end{equation}
while for the effective model the result is
\begin{equation}
P_3^{\eff}(\infty) = 
 \exp\bigg[-2\Gamma\int_{-\infty}^\infty \frac{\dot\theta^2 \left(\sin^4\varphi+\cos^4\varphi\right)}{\Omega_0^2+\dot\varphi^2}\mathrm{d}t'\bigg],
\end{equation}
where we have taken into account the fact that for the counterintuitive sequence, we have $P_0(-\infty)=P_1(-\infty)=1$ and $P_0(\infty)=P_3(\infty)$.

\begin{figure}
\includegraphics[width=0.45\textwidth, angle=0]{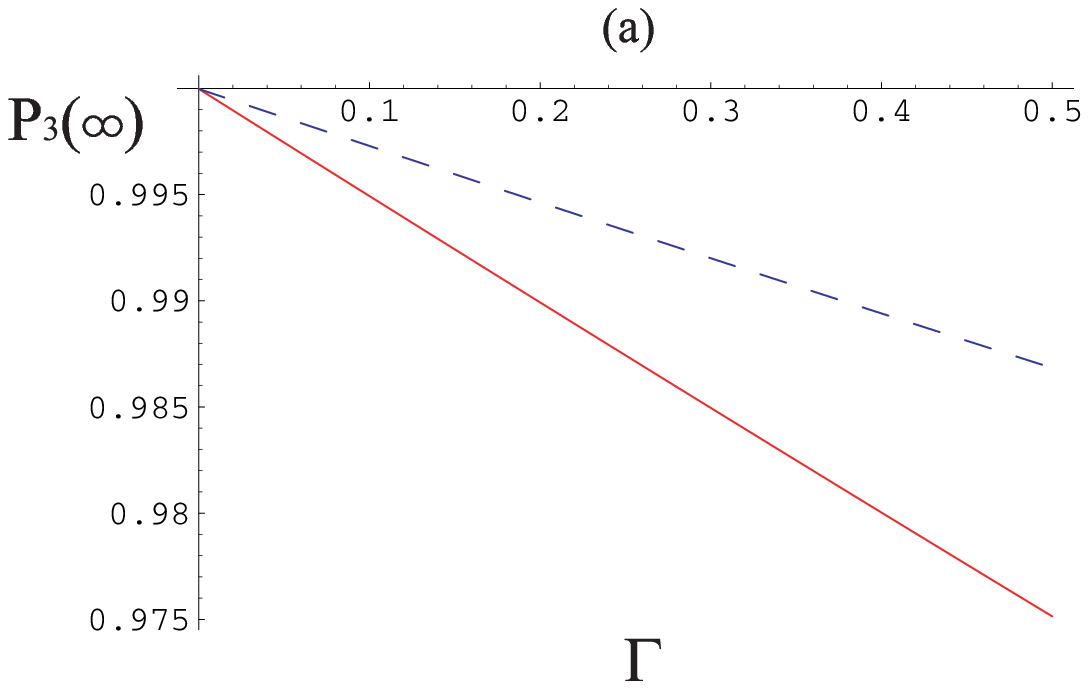}\\
\includegraphics[width=0.45\textwidth, angle=0]{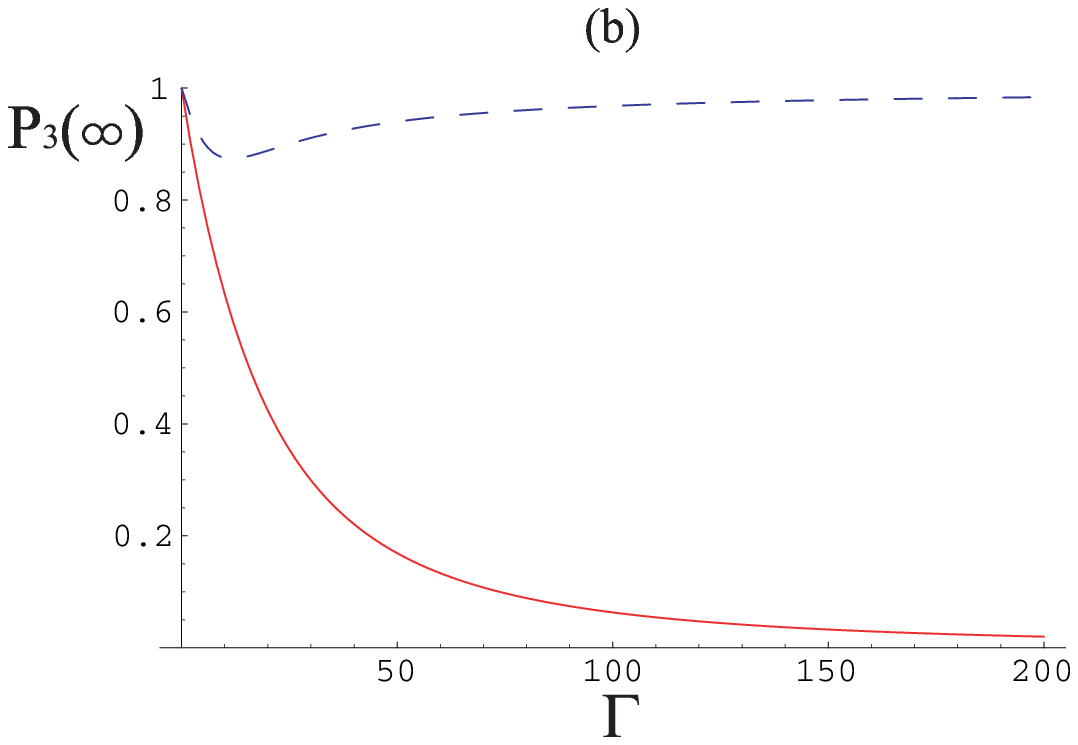}
\caption{(Color online) Post-pulse values of population $P_3$ for
the phenomenological (solid red line) and effective (dashed blue
line) model for the counterintuitive pulse sequence, as functions
of $\Gamma$ (in units of $T^{-1}$) for weak damping only (a) and
in a wide range (b). Relevant parameters: $\alpha T=10$, $\Delta
T=1$.} \label{fig:Phenom_vs_Effective-Asympt}
\end{figure}

Since $\sin^4\varphi+\cos^4\varphi\le 1$, it is evident that $P_3^{\eff}(\infty) > P_3^{\phen}(\infty)$.
Numerical results agree with this prediction, as shown in Fig.~ \ref{fig:Phenom_vs_Effective-Asympt}.

It can be shown that as $\alpha$ increases the agreement between
the results obtained in the adiabatic approximation and the
numerical results improves. Indeed, generally speaking, higher
$\alpha$ means higher Rabi frequencies between the states, which
put the system more under adiabatic conditions
\cite{ref:Vitanov1997b}. It is also worth nothing that the higher
$\alpha$ the closer to unity are the values of $P_3$ for
$\Gamma=0$. Again, this depends on the validity of the adiabatic
approximation, according to which the levels $\Ket{+}$ and
$\Ket{-}$ are not populated,
 so that the population transfer by $\Ket{0}$ is more and more efficient, without losses of population to the other two levels.

\subsection{Strong damping: numerical simulations}

Figure \ref{fig:Phenom_vs_Effective-Asympt}b shows the complete dependence of $P_3^{\eff}(\infty)$ and $P_3^{\phen}(\infty)$ on $\Gamma$ and, in particular, the strong damping limit.
While $P_3^{\phen}(\infty)$ decays as $\Gamma$ increases, it is well visible that $P_3^{\eff}(\infty)$ reaches higher values and in particular approaches $1$ for high values of $\Gamma$.
Such a behavior can be explained in terms of generalized quantum Zeno effect \cite{ref:QZE1,ref:QZE2}, meaning that the strong decay produces a separation of the Hilbert space into Zeno subspaces.
If $\Gamma$ is much larger than other quantities ($\Omega_0$, $\dot\varphi$, $\dot\theta$), we can split the
Hamiltonian $H_{\eff}$ as a sum of the \textit{unperturbed one}, which contains only terms proportional to $\Gamma$, and a
\textit{perturbation}, that is
\begin{align}
H_{\eff} =&\
 + \left[\begin{array}{cccc}
-i\Gamma\cos^2\varphi & 0 & 0\\
0 & 0 & 0\\
0 & 0 & -i\Gamma\sin^2\varphi\\
\end{array}\right]
 \notag\\
&+ \left[\begin{array}{cccc}
\Omega_0\cot\varphi & i\dot{\theta}\sin\varphi & i\dot{\varphi}\\
-i\dot{\theta}\sin\varphi & 0 & -i\dot{\theta}\cos\varphi\\
-i\dot{\varphi} & i\dot{\theta}\cos\varphi & -\Omega_0\tan\varphi\\
\end{array}\right]\,.
\end{align}
The three eigenvalues of the unperturbed (first) part of the
Hamiltonian are $\lambda=-i\Gamma\cos^2\varphi,
-i\Gamma\sin^2\varphi$ and 0, corresponding to the eigenstates
$\Ket{+}$, $\Ket{-}$, $\Ket{0}$. Since $\Gamma$ is very high, for
$\varphi\not=\pi/4$ the three eigenstates correspond to very
different eigenvalues and the presence of the
\textit{perturbation} only slightly changes the eigenvalues and
eigenstates of the total Hamiltonian, so that the three states
turn out to be essentially uncoupled throughout the process. On
the contrary, when $\varphi=0$ the states $\Ket{0}$ and $\Ket{-}$
could be coupled by the terms $\mp i\dot\theta\cos\varphi$.
However, it is easy to see that when $\varphi=0$, $\theta(t)$ is a
constant function and $\dot\theta=0$. Moreover, when
$\varphi=\pi/4$ the states $\Ket{+}$ and $\Ket{-}$ are coupled by
the terms $\pm i\dot\varphi$, but the relevant transitions are
unimportant since the only state with nonzero population is
$\Ket{0}$. Therefore, there are no transitions between the
adiabatic states, and $\Ket{0}$ can transfer the population from
level $\Ket{1}$ to level $\Ket{3}$ without decay since the
corresponding eigenvalue is essentially zero.

In the phenomenological model, we have
\begin{align}
H_{\phen} =&\  \left[ \begin{array}{cccc}
-i\Gamma\cos^2\varphi & 0 & \frac{i}{2}\Gamma\sin2\varphi\\
0 & 0 & 0\\
 \frac{i}{2}\Gamma\sin2\varphi & 0 & -i\Gamma\sin^2\varphi\\
\end{array}\right]
 \notag\\
&+\left[\begin{array}{cccc}
\Omega_0\cot\varphi & i\dot{\theta}\sin\varphi & i\dot{\varphi}\\
-i\dot{\theta}\sin\varphi & 0 & -i\dot{\theta}\cos\varphi\\
-i\dot{\varphi} & i\dot{\theta}\cos\varphi & -\Omega_0\tan\varphi\\
\end{array}\right]\,,
\end{align}
and the eigenvalues of the unperturbed (first) part of the Hamiltonian are $\lambda=0, 0$ and $-i\Gamma$, corresponding to the eigenstates $\Ket{0_1}=\Ket{0}$,
$\Ket{0_2}=\sin\varphi\Ket{+}+\cos\varphi\Ket{-}$ and $\Ket{-i\Gamma}=\cos\varphi\Ket{+}-\sin\varphi\Ket{-}$,
respectively. Since $\Gamma$ is very large, the perturbation does not couple the state $\Ket{-i\Gamma}$ to the doublet (the other
two states in the degenerate subspace), but the states of the doublet can be coupled by the perturbation. In fact, the
restriction of the perturbation to the doublet is
\begin{equation}
H_{\text{doublet}} = \left[\begin{array}{cc}
0 & -i\dot\theta\\
i\dot\theta & 0
\end{array}\right]\,,
\end{equation}
and since $\theta$ in the counterintuitive sequence varies from $0$ to $\pi/2$, there is complete population inversion in the doublet.
Therefore, since $\Ket{0_1}(-\infty)=\Ket{0}(-\infty)=\Ket{1}$, the final state of the system is $\Ket{0_2}(\infty)=\Ket{-}(\infty)=\Ket{1}$, and there is no population transfer to level $\Ket{3}$.
Summarizing, in the strong coupling for a counterintuitive pulse sequence, the effective model predicts a complete population transfer whereas
the phenomenological model predicts essentially no transfer.

Let us now turn to the intuitive pulse sequence. We have already
seen that for small values of $\Gamma$ the population transfer
from $\Ket{1}$ to $\Ket{3}$ is very low. This feature, which is
common to the two models, relies on very different physical
mechanisms. Let us start considering that in the intuitive
sequence, for the phenomenological model, at $t=-\infty$ we have
$\Ket{1}=\Ket{-}=\Ket{0_2}$, which, according to the previous
analysis, undergoes a complete transition toward state
$\Ket{0_1}$,  so that the final state is
$\Ket{0_1}(\infty)=\Ket{0}(\infty)=\Ket{1}$, and there is no
transfer to state $\Ket{3}$. Figure
\ref{fig:Evulution-VeryLargeGamm}a shows this behavior. In the
effective model, the system starts from $\Ket{-}(-\infty)=\Ket{1}$
and there are transitions between states $\Ket{+}$ and $\Ket{-}$
around $\varphi=\pi/4$; the entire subspace spanned by these two
states decays. Therefore, a complete loss of probability of the
system characterizes the dynamics in this case. Figure
\ref{fig:Evulution-VeryLargeGamm}b illustrates this feature.

\samepage{
\begin{figure}
\includegraphics[width=0.45\textwidth, angle=0]{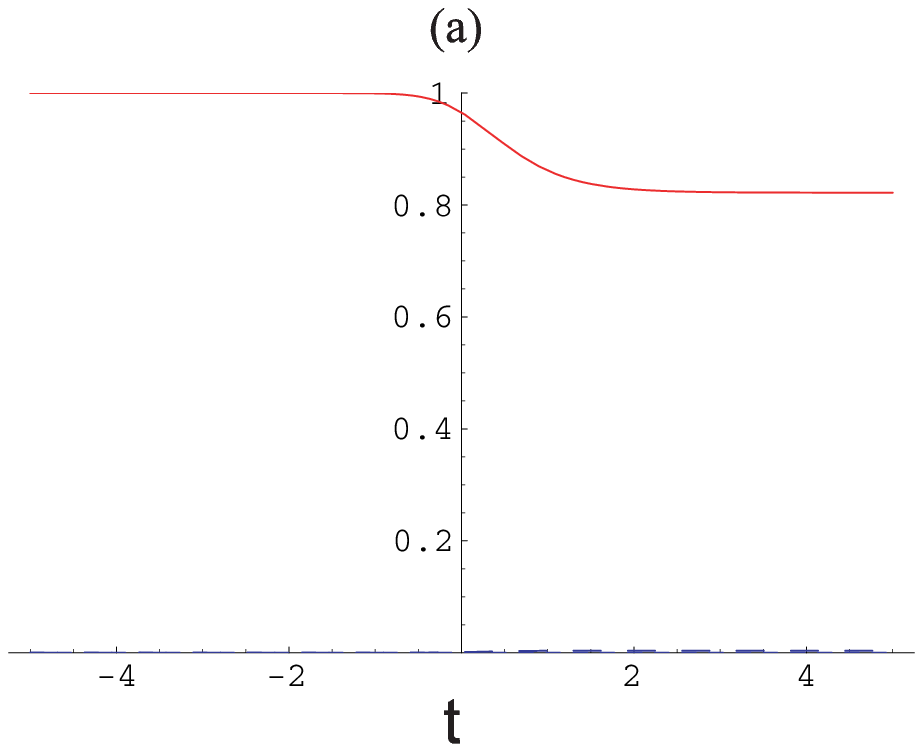}\\
\includegraphics[width=0.45\textwidth, angle=0]{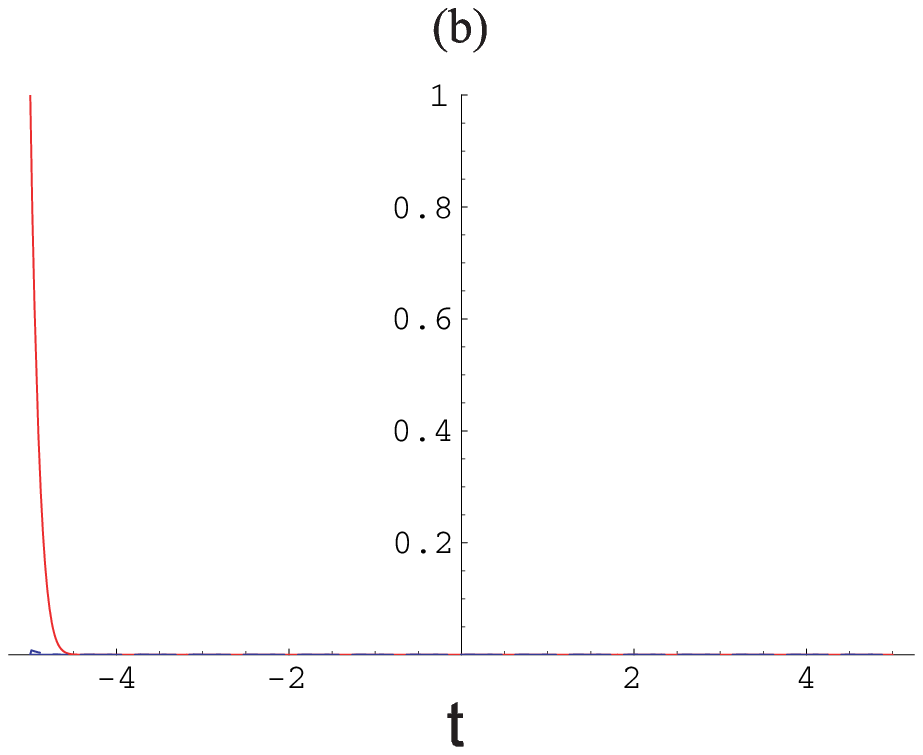}
\caption{(Color online) Evolution of the populations in the strong
damping limit, according to the effective model in the intuitive
sequence (a) and in the phenomenological model in the
counterintuitive sequence (b): $P_1$ as a solid (red) line, $P_2$
as a dotted (green) line and $P_3$ as a dashed (blue) line.
Populations $P_2$ and $P_3$ are essentially zero everywhere. Time
is measured in units of $T$. Relevant parameters are $\alpha
T=10$, $\Gamma T=500$, $\Delta T=1$.}
\label{fig:Evulution-VeryLargeGamm}
\end{figure}
}

\section{Conclusions}

In this paper we have shown that the phenomenological and
effective models for a laser-driven three-state $\Lambda$-system
under dissipative dynamics give different descriptions.

We have shown that the two relevant non-Hermitian Hamiltonian
models differ, in the adiabatic basis, in off-diagonal terms which
couple levels 1 and 3 and which are proportional to the decay
constant $\Gamma$. This suggests that the discrepancy between
predictions coming from the two models may increase as the decay
rate increases. Both analytical and numerical results confirm this
insight. Indeed, for weak damping the predictions from both models
are essentially the same, while in the strong damping limit the
difference is evident. Specifically, in the strong damping limit,
for the counterintuitive pulse sequence, there is a significant
difference in the values of the post-pulse population of the
target state: according to the phenomenological model there should
be no transfer from state $1$ to state $3$, while the effective
model predicts almost complete population transfer as $\Gamma$
increases. We note that, even though for the intuitive pulse
sequence the predictions for the post-pulse population of level
$3$ are the same for both models, the physical mechanisms that
lead to this result are very different for large $\Gamma$. In
fact, in the phenomenological model the population is kept in
state $1$, while in the effective model there is no final
population in state $3$ because all the states have undergone a
total decay.

The most important result in this paper is the prediction of a
complete population transfer for the counterintuitive sequence in
the strong damping limit; this result is of potential interest in
applications of STIRAP schemes in the manipulation of quantum
states.

\section*{Acknowledgements}

This work is supported by European Commission's projects EMALI and
FASTQUAST, and the Bulgarian Science Fund grants VU-F-205/06,
VU-I-301/07, and D002-90/08. Support from MIUR Project N.
II04C0E3F3 is also acknowledged.

\appendix

\section{Derivation of the Effective Model}\label{AppA}

In this appendix we derive the effective Hamiltonian model in
\eqref{eq:Effective_Hamiltonian} from the master equation
(\ref{eq:master_equation}).

Substituting \eqref{eq:rho_expansion} into
\eqref{eq:master_equation}, and assuming zero temperature
($\tilde{\gamma}$'s $= 0$) one obtains the following set of rate
equations in the adiabatic basis:
\begin{widetext}
\begin{subequations}
\begin{eqnarray}
&&\dot{\rho}_{00}=-\dot{\theta}\sin\varphi\,\rho_{+0}-\dot{\theta}\sin\varphi\,\rho_{0+}-\dot{\theta}\cos\varphi\,\rho_{-0}-\dot{\theta}\cos\varphi\,\rho_{0-}\,,\\
&&\dot{\rho}_{++}=-\gamma_{+}\,\rho_{++}+\dot{\varphi}\,\rho_{-+}+\dot{\varphi}\,\rho_{+-}+\dot{\theta}\sin\varphi\,\rho_{0+}+\dot{\theta}\sin\varphi\,\rho_{+0}\,,\\
&&\dot{\rho}_{--}=-\gamma_{-}\,\rho_{--}-\dot{\varphi}\,\rho_{-+}-\dot{\varphi}\,\rho_{+-}+\dot{\theta}\cos\varphi\,\rho_{0+}+\dot{\theta}\cos\varphi\,\rho_{+0}\,,\\
&&\dot{\rho}_{+0}=\left(-i\omega_{+0}-\frac{\gamma_+}{2}\right)\rho_{+0}+\dot{\theta}\sin\varphi\,\rho_{00}+\dot{\varphi}\rho_{-0}-\dot{\theta}\sin\varphi\,\rho_{++}
- \dot{\theta}\cos\varphi\,\rho_{+-}\,,\\
&&\dot{\rho}_{+-}=\left(-i\omega_{+-}-\frac{\gamma_+ +
\gamma_-}{2}\right)\rho_{+-}+\dot{\theta}\sin\varphi\,\rho_{0-}+\dot{\varphi}\rho_{--}+\dot{\theta}\cos\varphi\,\rho_{+0}
- \dot{\varphi},\rho_{++}\,,\\
&&\dot{\rho}_{0-}=\left(-i\omega_{0-}-\frac{\gamma_-}{2}\right)\rho_{0-}-\dot{\theta}\sin\varphi\,\rho_{+-}
-\dot{\theta}\cos\phi\,\rho_{--}-\dot{\varphi}\,\rho_{0+}+\dot{\theta}\cos\varphi\,\rho_{00}
- \dot{\varphi}\,\rho_{++}\,,
\end{eqnarray}
\end{subequations}
\end{widetext}
and the Hermitian conjugates of the last three equations.
Equations involving level $4$ are not shown, since they don't play
any role in the effective model. As a consequence the effective
model does not conserve the total probability. It is
straightforward to see that substituting the Hamiltonian model in
\eqref{eq:Effective_Hamiltonian} into \eqref{eq:Pseudo_Liouville},
one obtains exactly the same rate equations. Therefore we conclude
that $H_{\eff}$ represents the effective Hamiltonian of the
system.

\section{Strong damping in the bare basis}\label{AppB}

In this appendix we show how to treat the strong damping dynamics in the bare basis $\{\Ket{1},\Ket{2},\Ket{3}\}$.
In ref. \cite{ref:Vitanov1997} the starting point of the treatment is to give the phenomenological Hamiltonian:
\begin{equation}\label{eq:Hphbare}
H_{phen}^{bare}= \left[\begin{array}{ccc}
0 & \Omega_p(t) & 0  \\
 &  &  \\
\Omega_p(t) & \Delta-i\Gamma & \Omega_s(t)  \\
 &  &  \\
0 & \Omega_s(t) & 0 \\
\end{array}\right]\,,
\end{equation}
which, transformed to the adiabatic basis
$\{\Ket{+},\Ket{0},\Ket{-}\}$, gives Eq. (\ref{H
phenomenological}). Equation (\ref{eq:Hphbare}) clearly indicates
that in the strong damping limit state $\Ket{2}$ is well separated
from the other states,
 so that the coupling scheme does not allow to transfer population from state $\Ket{1}$ to state $\Ket{3}$.

In the master equation approach, the effective Hamiltonian in the
bare basis assumes a much more complicated form which does not
allow to separate the three bare states. Indeed, since the
effective Hamiltonian is related to the phenomenological one by
the relation
\begin{equation}
H_{\eff}=H_{\phen}-i\frac{\Gamma}{2}\sin 2\varphi
\left[\begin{array}{ccc}
0 & 0 & 1  \\
0 & 0 & 0  \\
1 & 0 & 0  \\
\end{array}\right]\,,
\end{equation}
we find that the inverse transformation from the adiabatic to the
bare basis gives:
\begin{widetext}
\begin{equation}
H_{\eff}^{\bare}= H_{\phen}^{\bare}-i\frac{\Gamma}{2}\sin 2\varphi
\left[\begin{array}{ccc}
2\sin\varphi\cos\varphi\sin^2\theta & \cos 2 \varphi\sin\theta & 2\sin\varphi\cos\varphi\sin\theta\cos\theta  \\
\cos 2 \varphi\sin\theta & -\sin 2\varphi & \cos 2 \varphi\cos\theta  \\
2\sin\varphi\cos\varphi\sin\theta\cos\theta & \cos 2 \varphi\cos\theta & 2\sin\varphi\cos\varphi\cos^2\theta  \\
\end{array}\right],
\end{equation}
\end{widetext}
which clearly shows the impossibility of separating state $\Ket{2}$ from the other ones.
Therefore the only basis in which the strong-damping dynamics can be easily explained is the adiabatic one.

\end{document}